\documentclass[twocolumn,epjc3]{svjour3}

\usepackage[T1]{fontenc}
\usepackage{mathptmx}
\usepackage{graphicx}
\usepackage{amsmath}
\usepackage{xspace}
\usepackage{flushend}
\usepackage[numbers,sort&compress]{natbib}
\usepackage[colorlinks,citecolor=blue,urlcolor=blue,linkcolor=blue]{hyperref}

\journalname{Eur. Phys. J. C}

\newcommand{\beqar}{\begin{eqnarray}}
\newcommand{\eeqar}{\end{eqnarray}}
\newcommand{\beq}{\begin{equation}}
\newcommand{\eeq}{\end{equation}}
\newcommand{\bit}{\begin{itemize}}
\newcommand{\eit}{\end{itemize}}

\def\refeq#1{\mbox{(\ref{#1})}}

\def\reffi#1{\mbox{Fig.~\ref{#1}}}
\def\reffis#1{\mbox{Figures~\ref{#1}}}
\def\refta#1{\mbox{Table~\ref{#1}}}

\def\ie{i.e.\ }

\newcommand{\ri}{\mathrm{i}}
\newcommand{\rF}{\mathrm{F}}
\newcommand{\rR}{\mathrm{R}}

\newcommand{\rT}{\mathrm{T}}
\newcommand{\rd}{\mathrm{d}}
\newcommand{\rs}{\mathrm{s}}
\newcommand{\rw}{\mathrm{w}}
\newcommand{\MW}{M_\mathrm{W}}
\newcommand{\MZ}{M_\mathrm{Z}}
\newcommand{\MH}{M_\mathrm{H}}
\newcommand{\mur}{\mu_{\rR}}
\newcommand{\muf}{\mu_{\rF}}

\newcommand{\MeV}{\mathrm{MeV}}
\newcommand{\GeV}{\mathrm{GeV}}

\newcommand{\fb}{\mathrm{fb}}
\newcommand{\alphaS}{\alpha_{\rs}}

\def\mathswitchr#1{\relax\ifmmode{\mathrm{#1}}\else$\mathrm{#1}$\fi}
\newcommand{\Pt}{\mathswitchr t}
\newcommand{\Pb}{\mathswitchr b}
\newcommand{\Pe}{\mathswitchr e}
\newcommand{\PW}{\mathswitchr W}
\newcommand{\PWp}{\mathswitchr W^+}
\newcommand{\PWm}{\mathswitchr W^-}
\newcommand{\PZ}{\mathswitchr Z}
\newcommand{\PH}{\mathswitchr H}
\newcommand{\Pg}{\mathswitchr g}

\newcommand{\Pp}{\mathswitchr p}
\newcommand{\nemn}{\nu_{\Pe} \Pe^+ \mu^- \bar{\nu}_{\mu}}
\newcommand{\ww}{\PW^+\PW^-}
\newcommand{\wt}{\PW\Pt}
\newcommand{\wmt}{\PWm\Pt}

\newcommand{\ttbar}{\Pt\bar\Pt}
\newcommand{\bbbar}{\Pb\bar\Pb}
\newcommand{\wwbb}{\ww\bbbar}
\newcommand{\muwwbb}{\mu_{\PW\PW\Pb\Pb}}

\newcommand{\nj}{N_{j}}
\newcommand{\nb}{N_{\Pb}}
\newcommand{\ppto}{\Pp\Pp\to}
\newcommand{\chib}{\chi_{\Pb}}
\newcommand{\chit}{\chi_{\Pt}}
\newcommand{\prob}{P_{\Pb}}
\newcommand{\prot}{P_{\Pt}}
\newcommand{\momb}{p_{\Pb}}
\newcommand{\momw}{p_{\PW}}
\newcommand{\ratio}{R}

\newcommand{\Mt}{m_{\Pt}}
\newcommand{\mt}{m_{\Pt}}
\newcommand{\Mb}{m_{\Pb}}

\newcommand{\GF}{{G_\mu}}

\newcommand{\ftw}{\mathrm{FtW}}
\newcommand{\LO}{\mathrm{LO}}
\newcommand{\NLO}{\mathrm{NLO}}

\newcommand{\Gt}{\Gamma_{\Pt}}
\newcommand{\GtLO}{\Gamma_{\Pt}^{\LO}}
\newcommand{\GtNLO}{\Gamma_{\Pt}^{\NLO}}

\newcommand{\ptjthr}{p^{\mathrm{thr}}_{\rT,\mathrm{jet}}}
\newcommand{\ptbjthr}{p^{\mathrm{thr}}_{\rT,\mathrm{bjet}}}

\newcommand{\Sherpa}{{\scshape Sherpa}\xspace}

\newcommand{\Amegic}{{\scshape Amegic++}\xspace}
\newcommand{\OpenLoops}{{\scshape OpenLoops}\xspace}
\newcommand{\Collier}{{\scshape Collier}\xspace}

\newcommand{\xs}[2]{#1}

 \newcommand{\gsim}
 {\;\raisebox{-.3em}{$\stackrel{\displaystyle >}{\sim}$}\;}

\begin{document}


\title{A unified NLO description of top-pair and associated Wt production}


\author{F.~Cascioli\thanksref{e1,addr1}
   \and S.~Kallweit\thanksref{e2,addr1}
   \and P.~Maierh\"ofer\thanksref{e3,addr1}
   \and S.~Pozzorini\thanksref{e4,addr1}}

\thankstext{e1}{e-mail: cascioli@physik.uzh.ch}
\thankstext{e2}{e-mail: kallweit@physik.uzh.ch}
\thankstext{e3}{e-mail: philipp@physik.uzh.ch}
\thankstext{e4}{e-mail: pozzorin@physik.uzh.ch}


\institute{Institut f\"ur Theoretische Physik,
           Universit\"at Z\"urich,
           8057 Z\"urich,
           Switzerland\label{addr1}}

\date{}

\maketitle

\begin{abstract}

We present an NLO simulation of $\PWp\PWm\bbbar$ production with massive
b-quarks at the LHC.  Off-shell and non-resonant
contributions associated with top-pair and \linebreak single-top channels and with
leptonic W-boson decays are consistently taken into account using the
complex-mass \linebreak scheme.  Thanks to the finite b-quark mass, $\wwbb$ predictions
can be extended to the whole b-quark phase space, thereby including
Wt-channel single-top contributions that originate from collinear $\Pg\to
\bbbar$ splittings in the four-flavour scheme.  This provides a consistent
NLO description of $\ttbar$ and Wt production and decay, including quantum
interference effects.  The simulation is also applicable to exclusive 0- and
1-jet bins, which is of great importance for Higgs-boson studies in the
$\PH\to \PWp\PWm$ channel and for any other analysis with large top
backgrounds and jet vetoes or jet bins.


\end{abstract}

\section{Introduction}

Top quarks are the heaviest known fundamental particles, and the precise
theoretical understanding of their production and decay mechanism, within
or beyond the Standard Model, has deep implications on countless aspects of
the LHC physics programme.  
At the LHC, top quarks are mainly produced as $\ttbar$ pairs and via
single-top production in the $t$-channel or in the associated Wt mode. 
At 8\,TeV these latter single-top channels amount to $40\%$ and $10\%$
of the $\ttbar$ cross section, respectively.  In spite of their smaller
cross sections, they play an important role as direct probes of top-quark
weak interactions and of their flavour structure.
The separation of top-production into individual top-pair and single-top
contributions poses non-trivial experimental and theoretical challenges,
which are mainly due to the similarity among the final states associated
with the various mechanisms of top-production and decay.  In particular,
the definition of $\ttbar$ and $\PW\Pt$ production involves notorious and
quite subtle theoretical issues~\cite{White:2009yt}.

In the five-flavour (5F) scheme, $\PW\Pt$ production proceeds via b-quark
induced partonic channels like $\Pg\Pb\to\PWm\PWp\Pb$, and the presence of a
single b-jet represents a clearly distinctive feature with respect to
$\wwbb$ final states associated with $\ttbar$ production.  However, beyond LO
this separation ceases to exist, since $\Pg\Pg\to\wwbb$ 
enters also the next-to-leading order
(NLO) corrections to $\PW\Pt$ production.  The resulting $\ttbar$
contamination represents a huge NLO correction, which jeopardises the
perturbative convergence of the $\PW\Pt$ cross section in the 5F scheme.  To
circumvent this problem within the 5F scheme, various approaches have been
proposed aimed at subtracting the contribution of a second top resonance in
$\Pp\Pp\to\PW\Pt+X$~\cite{White:2009yt}.  However, these prescriptions either break
gauge invariance or are not applicable to a realistic experimental setup. 
Moreover they neglect the quantum interference between top-pair and
single-top contributions. 

A theoretically more rigorous approach consists of \linebreak adopting the four-flavour
(4F) scheme, where initial-state \linebreak b-quarks result from gluons via explicit
$\Pg\to\bbbar$ splittings.  In this framework, the process $\ppto\wwbb+X$
provides a unified description of Wt and $\ttbar$
production~\cite{Kauer:2001sp}, and the presence of the $\ttbar$--Wt
interference at LO stabilises the perturbative expansion.  In the 4F scheme,
treating finite-top-width effects in the complex-mass
scheme~\cite{Denner:2005fg} ensures a consistent off-shell continuation of
top-quark propagators and allows one to include double-, single-, and
non-resonant contributions to $\ppto\wwbb+X$ with all relevant
interferences.
Moreover, the ill-defined separation of top-pair and Wt production can be
replaced by a gauge-invariant separation of $\ppto\wwbb$ into its
narrow-top-width limit, which corresponds to on-shell top-pair production and
decay, and a finite-width remainder that includes off-shell $\ttbar$ effects
as well as single-top and non-resonant contributions plus related
interferences.

The presence of four final-state particles and intermediate top-quark
resonances render the simulation of $\wwbb$ production quite
challenging beyond LO.  
First NLO calculations with massless b-quarks
have been presented in~\cite{Denner:2010jp,Denner:2012yc,Bevilacqua:2010qb}.
For $\wwbb$ production with two hard b-jets, apart from a few noticeable 
exceptions~\cite{Denner:2012yc}, 
most observables turn out to be completely dominated by the on-shell $\ttbar$ contribution. 
In phase-space regions with unresolved b-quarks,
the importance of off-shell and single-top contributions is expected to increase
quite substantially. However, due to the presence of collinear singularities, 
such regions are not accessible in the massless b-quark 
approximation of~\cite{Denner:2010jp,Denner:2012yc,Bevilacqua:2010qb}.
To fill this gap, in this paper we
present a complete NLO $\wwbb$ calculation including off-shell W-boson decays and massive
b-quarks in the 4F scheme.
A similar calculation has been presented very recently in~\cite{Frederix:2013gra}.
These simulations provide NLO accurate $\wwbb$ predictions in the full phase
space and allow one to investigate, for the first time, top-pair and
single-top production in presence of jet vetoes or jet bins, such as in the
case of the $\PH\to\PWp\PWm$ analysis.  An important advantage of NLO
$\wwbb$ predictions in the 4F scheme is that they provide a fully
differential NLO description of both final-state b-jets and a
correspondingly accurate modelling of jet vetoes, while in the 5F scheme a
similar level of accuracy for spectator b-quarks in Wt production
would require an NNLO calculation.

\section{Technical tools and ingredients of the calculation}
 
\begin{figure}
  \begin{center}
    \includegraphics{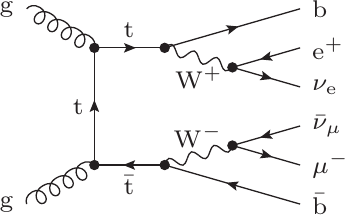}
    \includegraphics{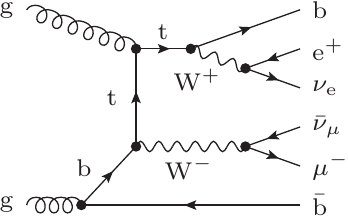}
  \end{center}
  \caption{Representative $\ttbar$-like  (left) and Wt-like (right) tree diagrams.}
  \label{fig:treegraphsA}
\end{figure}

\begin{figure}
  \begin{center}
    \includegraphics{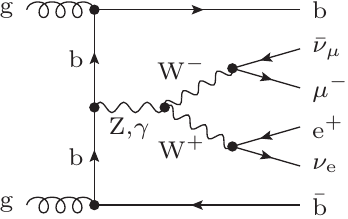}
    \includegraphics{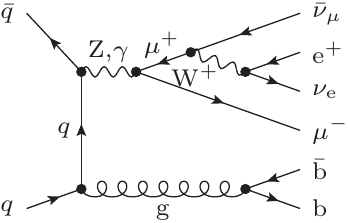}
  \end{center}
  \caption{Representative tree topologies without top resonances and with two (left) or only one (right) resonant W-boson.}
  \label{fig:treegraphsB}
\end{figure}

We will focus on NLO predictions for $\ppto\nemn\bbbar$, which comprises
$\ttbar$ production and decay in the opposite-flavour di-lepton channel.
For brevity we will denote this reaction as $\wwbb$ production, keeping in
mind that all off-shell and interference effects related to the $\nemn$
final state are consistently handled in the complex-mass
\linebreak scheme~\cite{Denner:2005fg}, where finite-width effects are systematically
absorbed in the imaginary part of the renormalised pole mass.
The complex-mass scheme is used also for the
off-shell continuation of top-quark resonances~\cite{Denner:2012yc}. 
Examples of tree diagrams involving two, one and no top-quark resonances are
illustrated in \reffis{fig:treegraphsA} and~\ref{fig:treegraphsB}.  The
second diagram in \reffi{fig:treegraphsA} is the 4F-scheme analogon of
$t$-channel $\Pg\Pb\to\Pt\PWm$ production in the 5F scheme, and
the initial-state $\Pg\to\bbbar$ splitting is related to the
b-quark parton distribution in 5F PDFs.  At NLO we include the full set
of tree, one-loop and real-emission diagrams that contribute to
$\nemn\bbbar$ production without applying any approximation.  In particular
non-resonant $Z/\gamma\to\nemn$ sub-topologies like in the second diagram of
\reffi{fig:treegraphsB} are included also in the virtual and real corrections.
The bottom- and top-quark masses are renormalised in the on-shell scheme, 
and their contributions are retained everywhere.

The entire calculation has been performed with highly flexible and
automated NLO programs, and the high complexity resulting from the presence
of multiple top- and W-resonances, as well as from the wide spectrum of
involved scales, render $\ppto\wwbb$ an excellent technical benchmark to
test the performance of the employed tools.
To evaluate tree, virtual, and real-emission amplitudes, we \linebreak employed
\OpenLoops~\cite{Cascioli:2011va},
a new one-loop generator that will become public
in the next future. The \OpenLoops program
is based on a novel numerical recursion, which is formulated in terms of loop-momentum
polynomials called ``open loops'' and allows for a fast evaluation of
scattering amplitudes with many external particles.  It uses the \linebreak \Collier
library~\cite{collier} for the numerically stable evaluation of tensor
integrals~\cite{Denner:2002ii,Denner:2005nn} and scalar
integrals~\cite{Denner:2010tr}.
Together with \cite{Cascioli:2013gfa,Cascioli:2013era}, the present study is one of the very
first applications of \OpenLoops.
Phase-space integration and infrared subtractions 
are performed with an in-house NLO Monte-Carlo framework~\cite{SKtool}, which 
is interfaced with \OpenLoops and provides full automation along the entire
chain of operations that are required
for NLO calculations. This tool is applicable to
any Standard-Model process at NLO QCD. Infrared singularities are handled with
dipole subtraction~\cite{Catani:1996vz,Catani:2002hc}, and since collinear $\Pg\to\bbbar$ 
splittings are regularised by the finite b-quark mass, corresponding subtraction terms
are not included.
The phase-space integrator is based on the
adaptive multi-channel technique~\cite{Kleiss:1994qy}
and implements dedicated channels for the dipole subtraction terms,
which improve the convergence, especially for \linebreak multi-resonance processes.
Multiple scale variations in a \linebreak single run are also supported.
This tool has been validated in several NLO processes and, in combination with \OpenLoops
and \Collier,
it is also applicable to NNLO calculations~\cite{Grazzini:2013bna}.
The correctness of the results is supported by various checks:
\OpenLoops has been validated against an independent in-house generator for more than hundred
partonic processes, including $\wwbb$ production with massless b-quarks and various processes
with massive heavy-quarks. For the process at hand we checked the cancellation of infrared and ultraviolet singularities.
The correctness of phase-space integration and dipole subtraction was tested
by means of a second calculation based on \OpenLoops in combination with
\Sherpa~\cite{Gleisberg:2007md,Gleisberg:2008ta} and \Amegic~\cite{Krauss:2001iv}.


\section{Input parameters, cuts and jet definition}

In the following, we present NLO results for $\wwbb$ production at the 8\,TeV LHC.
For the heavy-quark and  gauge-boson masses we use
\beqar
&\Mt=173.2~\GeV
,\quad
\Mb=4.75~\GeV
,&\nonumber\\
&\MZ=91.1876~\GeV
,\quad
\MW=80.385~\GeV.
\eeqar
The electroweak coupling is
derived from the Fermi constant,
$\GF=1.16637\times10^{-5}\GeV^{-2}$,
in the $\GF$-scheme,
\beq\label{eq:GFscheme}
\alpha=\frac{\sqrt{2}}{\pi}\GF\/\MW^2\left(1-\frac{\MW^2}{\MZ^2}\right).
\eeq
In the complex-mass scheme the electroweak mixing angle is evaluated as
\beq
\cos^2\theta_\rw = \frac{\MW^2-\ri\Gamma_\PW\MW}{\MZ^2-\ri\Gamma_\PZ\MZ},
\eeq
and for the widths we use the NLO QCD values
\beq
\label{eq:GammaWZ}
\Gamma_\PW=2.09530~\GeV 
,\qquad
\Gamma_\PZ=2.50479~\GeV
\eeq
everywhere, \ie for LO as well as for NLO matrix elements.
The Higgs-boson mass and width are set to
$\MH=126~\GeV$ and $\Gamma_\PH=4.21~\MeV$.
To guarantee consistent top-decay \linebreak branching fractions,
matrix elements and top-width input parameters must be taken at the same perturbative order.
For the LO and NLO top-quark widths we use the values
\beq\label{eq:GtfwW}
\GtLO=1.47451~\GeV
, \qquad
\GtNLO=1.34264~\GeV,
\eeq
which are computed with massive b-quarks
and off-shell W-bosons~\cite{Jezabek:1988iv}.
Consistently with the use of massive b-quarks we employ 4F parton
distributions.  Specifically, at NLO the LHApdf implementation of the
4F NNPDF2.3 parton distributions~\cite{Ball:2012cx} and the corresponding
running strong coupling are used.  More precisely, we use a
reference set\footnote{NNPDF23$\_$nlo$\_$FFN$\_$NF4$\_$as$\_$0118} that 
is obtained
from a variable-flavour set with $\alphaS^{(5)}(\MZ)=0.118$ via inverse
5F evolution down to $\mu_\rF=\Mb$ and subsequent upward evolution
with four active flavours.  Since the \linebreak NNPDF2.3 release does not include LO
parton distributions, for LO predictions we adopt the
NNPDF21$\_$lo$\_$nf4$\_$100 4F set, which corresponds to a
reference strong-coupling value $\alphaS^{(5)}(\MZ)=0.119$.
While the 4F running of $\alphaS$ misses heavy-quark-loop effects,
corresponding $\mathcal{O}(\alphaS)$ contributions are consistently included
in the virtual corrections via zero-\linebreak momentum subtraction of the top- and
bottom-quark loops in the renormalisation of $\alphaS$.

To investigate NLO corrections to top-pair and Wt production
we select events with two oppositely charged leptons, $\ell=\Pe^+,\mu^-$, with
\beqar\label{eq:LHCcuts}
p_{\mathrm{T},\ell}&>&20\,\GeV
, \quad
{|\eta_{\ell}|<2.5}
,\quad
{p_{\mathrm{T,miss}}>20\GeV},
\eeqar
where $p_{\mathrm{T,miss}}$  is obtained from the vector sum of the neutrinos' 
transverse momenta.
Final-state QCD partons, including b-quarks, are recombined into IR-safe
jets using the anti-$k_\rT$ algorithm~\cite{Cacciari:2008gp} with
jet-resolution parameter $R=0.4$. \linebreak Events are categorised according to
the total number, $\nj$, of jets with $p_\rT>30~\GeV$ and $|\eta|<2.5$ and
the number of b-jets, $\nb$, within the same acceptance region.
We classify as b-jet any jet involving at least a b-quark, which
includes also the case of collimated $\bbbar$ pairs resulting from the
splitting of energetic gluons.  In fixed-order calculations the implementation of this b-jet definition
is possible only in presence of massive b-quarks, while collimated $\bbbar$ pairs
must be handled as ``gluon-jets'' in the massless case.

\section{Scale choice for top-pair and single-top production}

In order to isolate off-shell and single-top effects associated with the
finite top-quark width ($\ftw$) we decompose the differential $\wwbb$ cross section as
\beq\label{nwa1}
\rd\sigma_{\wwbb}=
\rd\sigma_{\ttbar}
+\rd\sigma^{\ftw}_{\wwbb},
\eeq
where the ${\ttbar}$ term
represents  on-shell top-pair production
and decay in spin-correlated narrow-width approximation.
The $\ttbar$ contribution is obtained from
the numerical extrapolation of the full $\wwbb$ cross section in the
narrow-width limit~\cite{Denner:2012yc},
\beq\label{nwa2}
\rd\sigma_{\ttbar}=\lim_{\Gt\to 0}
\rd\tilde\sigma_{\wwbb}(\Gt),
\eeq
with
\beq\label{nwa3}
\rd\tilde\sigma_{\wwbb}(\Gt)=
\left(\frac{\Gt}{\Gt^{\mathrm{phys}}}\right)^2
\rd\sigma_{\wwbb}(\Gt),
\eeq
where the factor $({\Gt}/{\Gt^{\mathrm{phys}}})^2$
compensates the 1/$\Gt^2$ scaling of the cross section in such a
way that top-decay  branching fractions remain constant when $\Gt\to 0$.
By construction the $\rd \sigma^{\ftw}_{\wwbb}$ remainder in~\refeq{nwa1}
contains all finite-top-width effects, including off-shell $\ttbar$
production as well as single-top and non-resonant contributions.

As compared to $\wwbb$ production with two hard b-jets,
the fully inclusive case involves a much wider spectrum of scales, ranging from
$\Mb$ to $m_{\ttbar}$. This renders theoretical calculations significantly more involved.
In particular, given that the $\ttbar$ and Wt contributions to $\wwbb$ production
are characterised by very different scales, it is a priori not clear if 
a conventional QCD scale choice can ensure a perturbatively stable description
of both contributions. For $\ttbar$ production, a scale of the order of the
geometric average of the top-quark transverse energies,
\beq\label{eq:ttscale}
\mu_{\ttbar}^2=E_{\rT,\Pt} E_{\rT,\bar\Pt}
\quad\mbox{with}\quad
E^2_{\rT,i}=m_i^2+p^2_{\rT,i},
\eeq
is known to ensure a good perturbative convergence~\cite{Denner:2012yc}.
In the case of the single-top $\wmt$  contribution
one has to deal with two sub-processes: a collinear $g\to\bbbar$
initial-state splitting followed by $\Pg\Pb\to \PWm\Pt$ hard scattering.\footnote{The 
charge-conjugate channels are implicitly understood.}
The respective characteristic scales are
the bottom- and the top-quark transverse energies, $E_{\rT,\Pb}\ll E_{\rT,\Pt}$, and
a QCD scale of type
\beq\label{eq:twscale}
\mu_{\Pt\PWm}^2=E_{\rT,\Pt} E_{\rT,\bar\Pb}
\eeq
should represent an appropriate choice, since
\beq
\alphaS^2(\mu^2_{\Pt\PWm}) \simeq \alphaS(E^2_{\rT,\Pt})
\alphaS(E^2_{\rT,\bar\Pb})
\eeq
guarantees that the $\alphaS$ factor associated with
the collinear $\Pg\to\bbbar$ splitting is effectively evaluated at the scale
$E_{\rT,\Pb}$, similarly as in the resummation of initial-state
b-quark emissions in the evolution of 5F PDFs.
Vice versa,   using a global QCD scale of the order $\Mt$
might underestimate the single-top component 
of $\ppto\wwbb$ by up to a factor \linebreak
$\alphaS(\Mb)/\alphaS(\Mt)\sim 2$ at LO.
This would be compensated by $\ln(\Mb)$-enhanced higher-order corrections,
resulting in a \linebreak poor perturbative convergence.
For an accurate description of the single-top contribution, the above considerations
motivate a dynamic QCD scale that interpolates between
\refeq{eq:ttscale}  and \refeq{eq:twscale} in $\ttbar$- and Wt-dominated regions,
respectively. Such a scale can be defined as
\beq\label{eq:wwbbscale1}
\muwwbb^2=
\mu_{\PWp\Pb}\,
\mu_{\PWm\bar\Pb},
\eeq
with
\beq\label{eq:wwbbscale2}
\mu_{\PW\Pb}
=
\prob
(\momw,\momb)\,
E_{\rT,\Pb}+
\prot(\momw,\momb)\,
E_{\rT,\Pt},
\eeq
where $\PW\Pb$ represents either $\PWp\Pb$ or $\PWm\bar\Pb$, and the
functions $\prob$ and $\prot=1-\prob$ describe
the probability that
the b-quark of a given $\PW\Pb$ pair
arises from an initial-state $\Pg\to\bbbar$ splitting or from a
$\Pt\to\PW\Pb$ decay, respectively.
Their approximate functional form can be obtained from the leading matrix-element
singularities associated with the $\Pg\to\bbbar$ and
$\Pt\to\PW\Pb$  sub-processes,\footnote{
The $\chib$ and $\chit$ distributions are defined as dimensionless functions 
by introducing $\Mt$-terms in the
numerator. This convention is however irrelevant, since 
the probabilities resulting from \refeq{eq:wwbbscale4} and \refeq{eq:wwbbscale5}
are independent of the normalisation of $\chib$ and $\chit$.}
\beq\label{eq:wwbbscale3}
\chib
= \frac{\Mt^2}{E^2_{\rT,\Pb}},
\quad
\chit 
= \frac{\Mt^4}{[(\momw+\momb)^2-\Mt^2]^2+\Gt^2\Mt^2},
\eeq
by requiring that $\prob/\prot\propto\chib/\chit$. This
yields
\beq\label{eq:wwbbscale4}
\prob = 1-\prot =
\frac{\chib}{\chib+\ratio\chit}.
\eeq
The constant $\ratio$
can be derived from the condition
 \beq\label{eq:wwbbscale5}
\int\rd \sigma^{\ftw}_{\wwbb}
=
\int \rd\Phi \left[1-P_{\Pt}(\Phi) P_{\bar\Pt}(\Phi)\right]
\frac{\rd\sigma_{\wwbb}}{\rd\Phi},
\eeq
\ie by requiring that finite-top-width corrections to the inclusive $\wwbb$ cross
section correspond to the contribution from non-$\ttbar$ events according to the
probability distributions $\prob$ and $\prot$.\footnote{Here we assume that finite-top-width effects
are dominated by non-$\ttbar$ contributions. Note also that the finite-top-width term on the left-hand side of
\refeq{eq:wwbbscale5} must be extracted through $\Gt\to 0 $ extrapolation by keeping $\Gt$ and $\ratio$ fixed in
\refeq{eq:wwbbscale3}--\refeq{eq:wwbbscale4}.}
The tuning of $\ratio$ is performed in LO approximation
on the fully inclusive level and yields $\ratio=7.96$.
At NLO, the kinematic quantities that enter $\muwwbb$
are defined in terms of b- and $\bar\Pb$-jet momenta that
are constructed with a modified jet algorithm
where $\bbbar$ pairs are not clustered and light partons with $|\eta|>4.5$
are excluded from the recombination  procedure.
The latter prescription guarantees the collinear safety
of the reconstructed top mass, $(p_{\PW}+p_{\Pb})^2$,
with respect to collinear light-parton emission from the initial state.
In the reconstruction of the top and anti-top masses $(p_{\PW}+p_{\Pb})^2$ that enter
\refeq{eq:wwbbscale3}, remaining hard jets are clustered with the $\Pt$- or $\bar\Pt$- system
if the resulting invariant mass turns out to be closer to $\Mt$.
Top-jet clusterings are applied only if they yield $P_\Pt>0.5$. If that holds for $\Pt$- and $\bar\Pt$- system, the clustering to maximise the $\ttbar$ probability, $P_{\Pt}P_{\bar\Pt}$, is chosen.

\section{Predictions for the LHC at 8\,TeV}

In the following we present predictions for $\ppto\wwbb$ at 8\,TeV
in presence of the leptonic cuts \refeq{eq:LHCcuts}.  If not stated
otherwise, the renormalisation and factorisation scales are set to
\beq\label{eq:RFscales} \mu_{\rR,\rF}=\xi_{\rR,\rF}\mu_0
\quad\mbox{with}\quad \mu_0={\muwwbb}, \eeq where $\xi_{\rR}=\xi_{\rF}=1$
corresponds to the default scale choice.  Theoretical uncertainties are assessed 
by applying the scale variations $(\xi_\rR,\xi_\rF)=(2,2)$, $(2,1)$, $(1,2)$, $(1,0.5)$, $(0.5,1)$,
$(0.5,0.5)$.

\begin{figure}
\begin{center}
\hspace*{-15mm}
{\includegraphics[width=.45\textwidth]{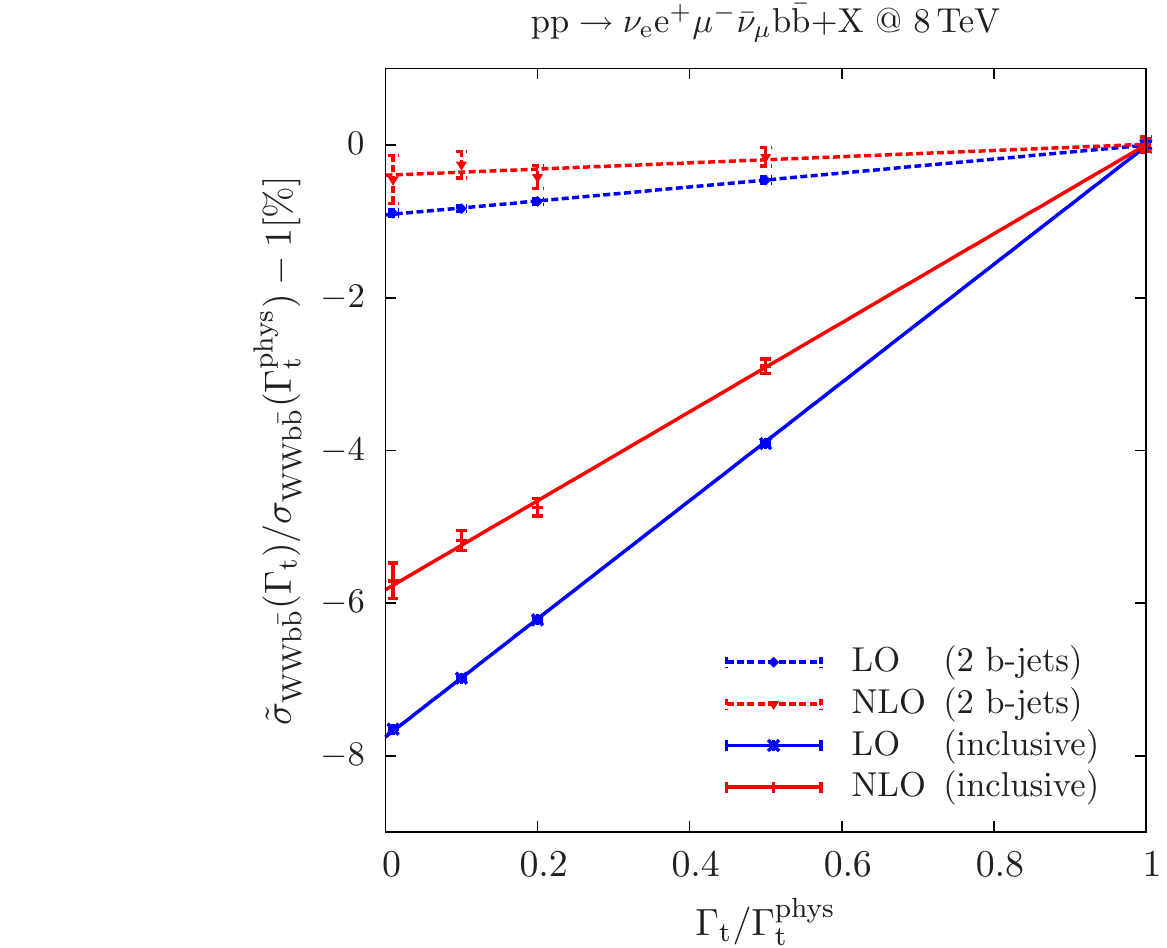}}
\end{center}
\caption{Numerical extrapolation of the LO and NLO $\wwbb$ cross section with leptonic cuts
in the narrow-top-width limit, $\Gt\to 0$. Results are shown as relative deviations (in percent)
with respect to the $\wwbb$ cross section with $\Gt=\Gt^{\mathrm{phys}}$. 
Results with inclusive jet emission are compared to a $\ttbar$-signal analysis
with two b-jets.
}
\label{fig:nwext}
\vspace{-1.mm}
\end{figure}

Figure~\ref{fig:nwext} illustrates the extrapolation of the $\wwbb$ cross section
in the narrow-top-width limit \refeq{nwa1}--\refeq{nwa2}.  
The results are well consistent---at the
few-permil level---with the expected linear convergence of the NLO
cross section in the $\Gt\to 0$ limit. This provides a non-trivial check of the consistency of
the calculation, since the narrow-width limit involves delicate
cancellations of logarithmic singularities that arise from virtual and real
soft-gluon corrections to the resonant top-quark propagators.
Finite-width effects turn out to be at the sub-percent level if
one requires the presence of two b-jets, like in a typical $\ttbar$-signal
analysis. For
the total cross section they are instead clearly more important.  Their net effect,
which results from the interplay of
negative off-shell corrections and positive single-top contributions,
amounts to about +6\%(8\%) at NLO(LO).

\begin{table}[h]
\setlength{\tabcolsep}{1.2ex}%
\caption{
LO and NLO predictions for $\ppto\wwbb$ at 8\,TeV
with scale variations and corrections, $K=\sigma_{\NLO}/\sigma_{\LO}$,
for different scale choices:
total cross section with leptonic cuts
and partial contributions with 0,1 and $\ge 2$ jets.
Full $\wwbb$ predictions ($\sigma$) are compared to finite-top-width
contributions ($\sigma^{\ftw}$). 
}
\label{tab:XSjet}
\begin{center}
\begin{tabular}{llllll}\hline
&  $\mu_0$
&  $\sigma[\fb]$
&  $\sigma_{0}[\fb]$
&  $\sigma_{1}[\fb]$
&  $\sigma_{2^+}[\fb]$
\\[.7mm] \hline
{$\LO$}
&  $\muwwbb$
&  $\xs{1232}{(2)}^{+34\%}_{-24\%}$       
&  $\xs{   37}{(2)}^{+38\%}_{-25\%}$      
&  $\xs{  367}{(2)}^{+36\%}_{-24\%}$      
&  $\xs{  828}{(2)}^{+33\%}_{-23\%}$      
\\
{$\NLO$}
&  $\muwwbb$
&  $\xs{1777}{(2)}^{+10\%}_{-12\%}$        
&  $\xs{   41}{(2)}^{+3\%}_{-8\%}$        
&  $\xs{  377}{(2)}^{+1\%}_{-6\%}$        
&  $\xs{ 1359}{(2)}^{+14\%}_{-14\%}$      
\\
{$K$}
&  $\muwwbb$
&  1.44
&  1.09
&  1.03
&  1.64

\\[.5mm] \hline
{$\LO$}
&  $\mt$
&  $\xs{1317}{(2)}^{+35\%}_{-24\%}$       
&  $\xs{   35}{(2)}^{+37\%}_{-25\%}$      
&  $\xs{  373}{(2)}^{+36\%}_{-24\%}$      
&  $\xs{  909}{(2)}^{+35\%}_{-24\%}$      
\\
{$\NLO$}
&  $\mt$
&  $\xs{1817}{(2)}^{+8\%}_{-11\%}$        
&  $\xs{   40}{(2)}^{+4\%}_{-8\%}$        
&  $\xs{  372}{(2)}^{+1\%}_{-8\%}$        
&  $\xs{ 1405}{(2)}^{+13\%}_{-13\%}$      
\\
{$K$}
&  $\mt$
&  1.38
&  1.14
&  1.00
&  1.55

\\\hline
&  $\mu_0$
&  ${\sigma^{\ftw}}[\fb]$
&  ${\sigma^{\ftw}_{0}}[\fb]$
&  ${\sigma^{\ftw}_{1}}[\fb]$
&  ${\sigma^{\ftw}_{2^+}}[\fb]$
\\[0.7mm] \hline
{$\LO$}
&  $\muwwbb$
&  $\xs{91}{(2)}^{+41\%}_{-27\%}$         
&  $\xs{   13}{(2)}^{+42\%}_{-27\%}$      
&  $\xs{   71}{(2)}^{+40\%}_{-27\%}$      
&  $\xs{    7}{(2)}^{+45\%}_{-29\%}$      
\\
{$\NLO$}
&  $\muwwbb$
&  $\xs{107}{(2)}^{+6\%}_{-11\%}$         
&  $\xs{   13}{(2)}^{+1\%}_{-7\%}$        
&  $\xs{   61}{(2)}^{+2\%}_{-16\%}$       
&  $\xs{   33}{(2)}^{+51\%}_{-31\%}$      
\\
{$K$}
&  $\muwwbb$
&  1.18
&  0.99
&  0.86
&  4.70

\\[.5mm] \hline
{$\LO$}
&  $\mt$
&  $\xs{63}{(2)}^{+36\%}_{-25\%}$         
&  $\xs{    8}{(2)}^{+36\%}_{-25\%}$      
&  $\xs{   49}{(2)}^{+36\%}_{-24\%}$      
&  $\xs{    6}{(2)}^{+46\%}_{-29\%}$      
\\
{$\NLO$}
&  $\mt$
&  $\xs{100}{(2)}^{+17\%}_{-16\%}$         
&  $\xs{   13}{(2)}^{+14\%}_{-14\%}$      
&  $\xs{   65}{(2)}^{+9\%}_{-12\%}$       
&  $\xs{   23}{(2)}^{+42\%}_{-28\%}$      
\\
{$K$}
&  $\mt$
&  1.58
&  1.47
&  1.32
&  3.89

\\\hline


\end{tabular}
\end{center}
\end{table}

Predictions for the integrated cross section and in exclusive jet bins are
listed in \refta{tab:XSjet}. To assess the influence of the scale choice,
results based on $\mu_0=\muwwbb$ are compared to the case of the
conventional scale $\mu_0=\Mt$.  For the total cross section
we find positive corrections of about 40\%.\footnote{
We note that these results are not directly comparable 
to those of~\cite{Denner:2012yc}, 
which reports a significantly smaller $K$-factor. In particular, 
while we apply the same cuts on leptons, missing energy and jets,
here we do not restrict ourselves to the case of two b-jets, 
we adopt a smaller jet-resolution parameter and a different QCD scale choice.
Moreover we employ a 4F PDF set, which implies an enhancement of the gluon density 
due to the absence of $\Pg\to\bbbar$ splittings in the PDF evolution. 
The LO PDF sets used in~\cite{Denner:2012yc} 
and in the present study feature also significantly different values of $\alphaS$, 
which influences LO results and $K$-factors. Finally, in addition to uniform scale 
variations considered in~\cite{Denner:2012yc}, here also
independent $\mur$ and $\muf$ variations are taken into account.}
Scale uncertainties decrease from about 30\% at LO to 10\% at NLO, and the
differences between the two scale choices are consistent within scale
variations.  
The last three columns of \refta{tab:XSjet} display jet cross
sections in bins with $\nj=0,1$ and $\nj\ge 2$ jets, where $\nj$ refers to
the total number of b-jets and light jets.  The different bins receive quite
different corrections, and the relative weight of the individual bins in
percent changes from 3:30:67 at LO to 2:21:76 at NLO.  This indicates that a
significant fraction of the 0- and 1-jet bin cross sections migrates to the
inclusive 2-jet bin. We attribute this feature 
to the rather high probability of light-jet emissions with $p_\rT\gsim 30\,\GeV$.  
While NLO scale uncertainties turn out to be fairly small in
all jet bins, matching to the parton
shower is certainly important for a more reliable description of such
radiative processes.  Comparing the two scale choices, also in jet bins we do not
observe any dramatic difference: absolute LO and NLO results are well
consistent within scale variations, and also 
K-factors and scale variations themselves turn out to be quite similar.
 
Finite-top-width (FtW) contributions are shown in the \linebreak lower part of
\refta{tab:XSjet}.  For what concerns the total $\wwbb$ cross section their
impact is around $6\%$, and the scale $\muwwbb$ guarantees a good
perturbative convergence: FtW contributions receive only minor NLO
corrections, and the residual scale dependence is about 10\%, while setting
$\mu_0=\Mt$ yields larger corrections and scale uncertainties.
As compared to complete $\wwbb$ predictions, FtW contributions are distributed
in a completely different way among jet bins.  The relative weight in percent
of the 0-, 1- and 2-jet bins is 14:78:8 at LO and 12:57:31 at NLO. 
These results suggest that FtW effects are dominated by a single-top $\wt$
component, which is concentrated in the 1-jet bin at LO and tends to migrate
to the 2-jet bin due to light-jet emissions at NLO.  The fact that the FtW part
of the 2-jet bin features a 40--50\% NLO uncertainty is irrelevant, since this
contribution represents less than 3\% of the complete cross section in the
2-jet bin.  In the 0- and 1-jet bins, whose FtW components amount to 32\%
and 16\%, respectively, NLO scale uncertainties are as small as 10\% or so.

\begin{table}
\setlength{\tabcolsep}{1.2ex}%
\caption{Full $\wwbb$ predictions and finite-top-width contributions 
for bins with 0,1 and $\ge2$ b-jets.
Same conventions as in~\refta{tab:XSjet}.}
\label{tab:XSbjet}
\begin{center}
\begin{tabular}{llllll}\hline
&  $\mu_0$
&  $\sigma[\fb]$
&  $\sigma_{0}[\fb]$
&  $\sigma_{1}[\fb]$
&  $\sigma_{2^+}[\fb]$
\\[.7mm] \hline
{$\LO$}
&  $\muwwbb$
&  $\xs{1232}{(2)}^{+34\%}_{-24\%}$       
&  $\xs{   37}{(2)}^{+38\%}_{-25\%}$      
&  $\xs{  367}{(2)}^{+36\%}_{-24\%}$      
&  $\xs{  828}{(2)}^{+33\%}_{-23\%}$      
\\
{$\NLO$}
&  $\muwwbb$
&  $\xs{1777}{(2)}^{+10\%}_{-12\%}$        
&  $\xs{   65}{(2)}^{+20\%}_{-17\%}$      
&  $\xs{  571}{(2)}^{+14\%}_{-14\%}$      
&  $\xs{ 1140}{(2)}^{+7\%}_{-10\%}$       
\\
{$K$}
&  $\muwwbb$
&  1.44
&  1.73
&  1.56
&  1.38

\\[.5mm] \hline
{$\LO$}
&  $\mt$
&  $\xs{1317}{(2)}^{+35\%}_{-24\%}$       
&  $\xs{   35}{(2)}^{+37\%}_{-25\%}$      
&  $\xs{  373}{(2)}^{+36\%}_{-24\%}$      
&  $\xs{  909}{(2)}^{+35\%}_{-24\%}$      
\\
{$\NLO$}
&  $\mt$
&  $\xs{1817}{(2)}^{+8\%}_{-11\%}$        
&  $\xs{   63}{(2)}^{+20\%}_{-17\%}$      
&  $\xs{  584}{(2)}^{+14\%}_{-14\%}$      
&  $\xs{ 1170}{(2)}^{+5\%}_{-9\%}$        
\\
{$K$}
&  $\mt$
&  1.38
&  1.80
&  1.56
&  1.29

\\\hline
&  $\mu_0$
&  ${\sigma^{\ftw}}[\fb]$
&  ${\sigma^{\ftw}_{0}}[\fb]$
&  ${\sigma^{\ftw}_{1}}[\fb]$
&  ${\sigma^{\ftw}_{2^+}}[\fb]$
\\[0.7mm] \hline
{$\LO$}
&  $\muwwbb$
&  $\xs{91}{(2)}^{+41\%}_{-27\%}$         
&  $\xs{   13}{(2)}^{+42\%}_{-27\%}$      
&  $\xs{   71}{(2)}^{+40\%}_{-27\%}$      
&  $\xs{    7}{(2)}^{+45\%}_{-29\%}$      
\\
{$\NLO$}
&  $\muwwbb$
&  $\xs{107}{(2)}^{+6\%}_{-11\%}$         
&  $\xs{   20}{(2)}^{+18\%}_{-17\%}$      
&  $\xs{   82}{(2)}^{+4\%}_{-10\%}$       
&  $\xs{    5}{(2)}^{+2\%}_{-10\%}$       
\\
{$K$}
&  $\muwwbb$
&  1.18
&  1.49
&  1.16
&  0.77

\\[.5mm] \hline
{$\LO$}
&  $\mt$
&  $\xs{63}{(2)}^{+36\%}_{-25\%}$         
&  $\xs{    8}{(2)}^{+36\%}_{-25\%}$      
&  $\xs{   49}{(2)}^{+36\%}_{-24\%}$      
&  $\xs{    6}{(2)}^{+46\%}_{-29\%}$      
\\
{$\NLO$}
&  $\mt$
&  $\xs{100}{(2)}^{+17\%}_{-16\%}$         
&  $\xs{   16}{(2)}^{+22\%}_{-18\%}$      
&  $\xs{   77}{(2)}^{+16\%}_{-15\%}$      
&  $\xs{    6}{(2)}^{+12\%}_{-16\%}$      
\\
{$K$}
&  $\mt$
&  1.58
&  1.89
&  1.58
&  1.10

\\\hline


\end{tabular}
\end{center}
\end{table}

In \refta{tab:XSbjet} we report analogous results for the $\wwbb$ cross
section and its FtW contribution in b-jet bins.  As compared to the case of
generic jets, we observe that $\wwbb$ K-factors feature a less pronounced 
dependence on the b-jet multiplicity
if the $\muwwbb$ scale is used. 
This is due to the fact that NLO emissions consist of light jets and 
are thus less likely to induce bin migrations in the case of b-jet bins.  
Scale uncertainties at NLO
are at the 20\%, 15\% and 10\% level in the bins with 0, 1, and $\ge 2$ b-jets, 
respectively.  Finite-top-width contributions turn out to be even
more stable than full $\wwbb$ results with the scale $\muwwbb$, while the
scale $\Mt$ tends to give larger uncertainties.  Using the $\muwwbb$ scale,
FtW effects in the 0-, 1-, and 2-b-jet bins turn out to be 31, 14 and 0.4
percent of the respective $\wwbb$ cross sections at NLO.  Employing
$\mu_0=\Mt$ these percentages become 25, 13 and 0.5, respectively.  In
general, jet- and b-jet-bin results indicate that the conventional scale
$\mu_0=\Mt$ yields a similarly good perturbative convergence as
$\mu_0=\muwwbb$.  However, it is a priori not clear 
if this holds also for more exclusive observables.  For what concerns theoretical
uncertainties in jet and b-jet bins, we checked that NLO scale variations remain
similarly small as in Tables~\ref{tab:XSjet}--\ref{tab:XSbjet} if the
jet-rapidity acceptance is increased up to $|\eta|<4.5$.

\begin{figure*}
\begin{center}
{\includegraphics[width=.45\textwidth]{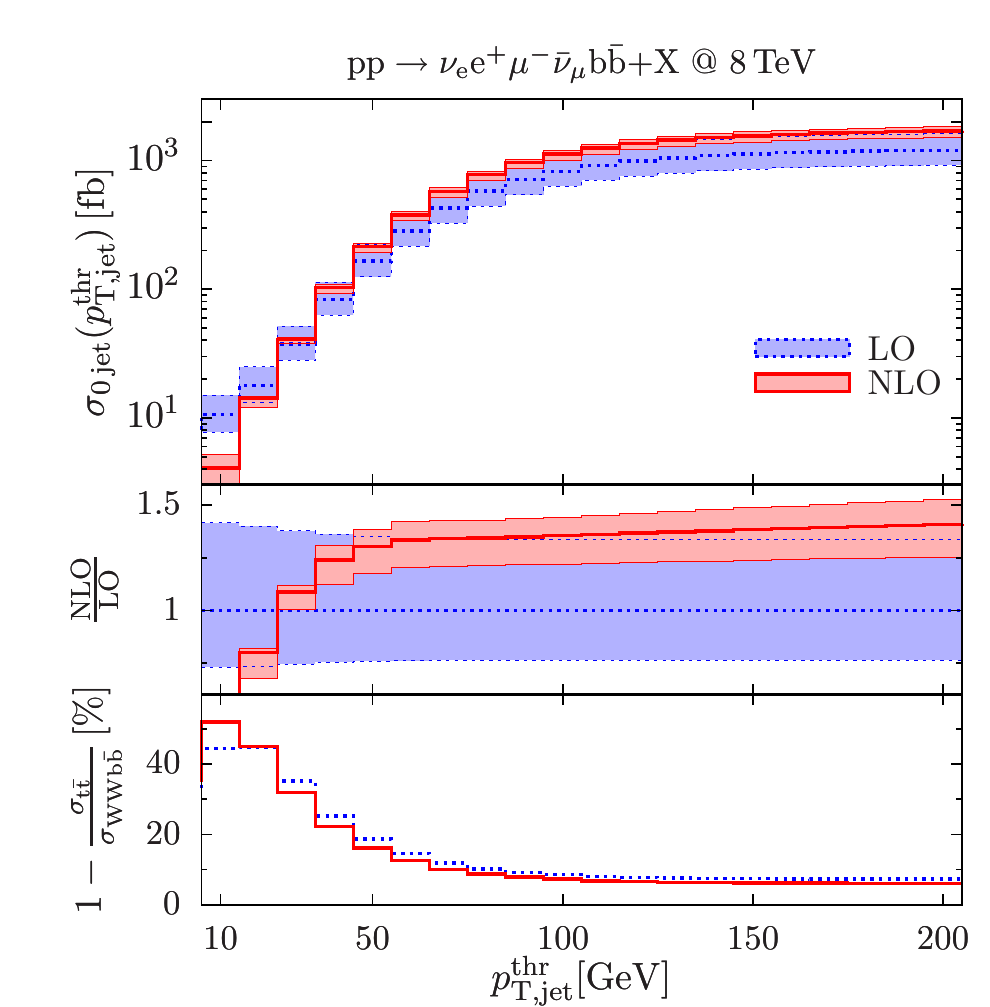}}
\hspace{10mm}
{\includegraphics[width=.45\textwidth]{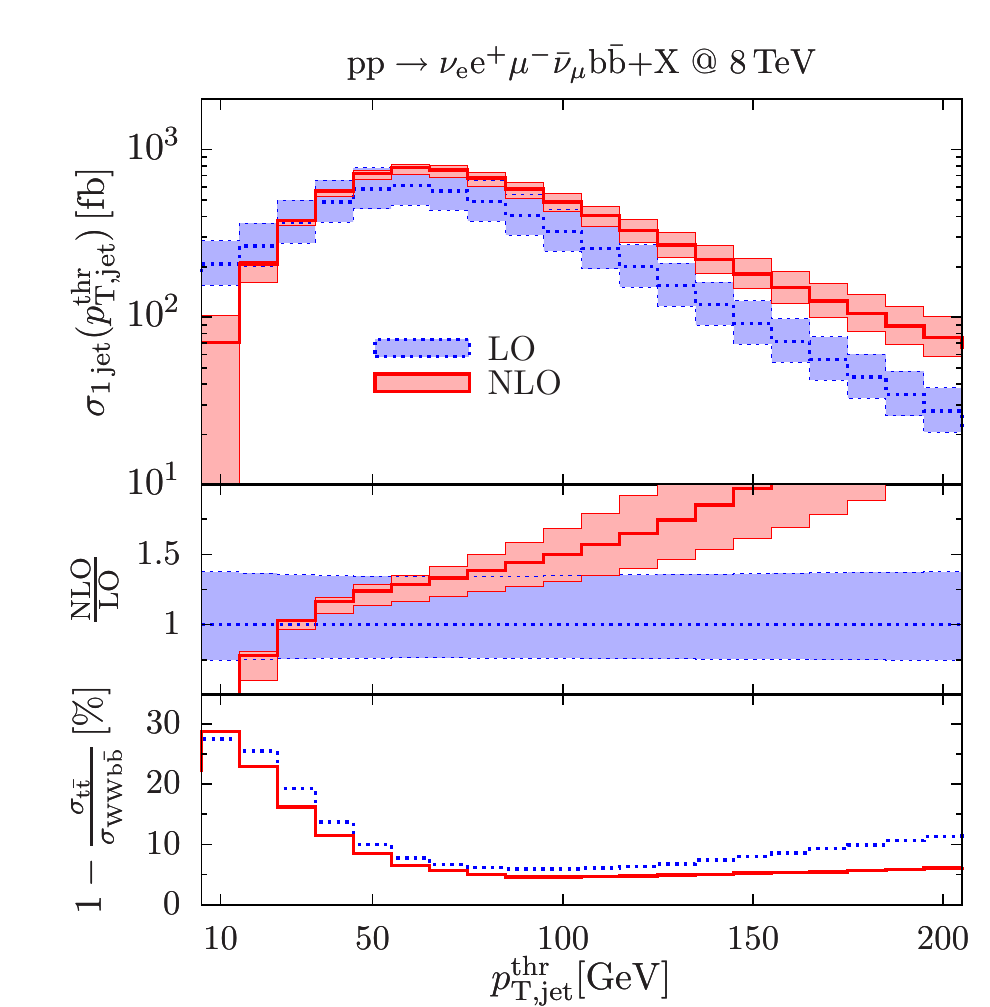}}
\end{center}
\caption{
LO and NLO $\wwbb$ cross sections in the exclusive bins with $\nj=0$ (left) and $\nj=1$ (right) 
jets as functions of the jet-$p_\rT$ threshold, $\ptjthr$. The middle of each bin corresponds to 
the actual value of $\ptjthr$. The central and lower frames show the $K$-factor and the relative 
impact in percent of finite-top-width contributions.
Where depicted, bands correspond to independent scale variations of $\mu_{\rR,\rF}$ by a 
factor of two around the central scale $\muwwbb$, not taking into account antipodal variations.}
\label{fig:ptjet}
\vspace{-1.mm}
\end{figure*}

To illustrate jet-veto and jet-binning effects in more detail, in
\reffi{fig:ptjet} we plot the integrated $\wwbb$ cross section in exclusive
bins with $\nj=0$ and $\nj=1$ jets versus the $p_\rT$-threshold that defines
jets.  The 0-jet bin corresponds to the integrated cross section in presence
of a jet veto, $p_{\rT,\mathrm{jet}}<\ptjthr$.  At large $\ptjthr$ the
$K$-factor and the FtW contributions converge quite smoothly towards their
inclusive limit.  In contrast, the region of small transverse momentum
features a very pronounced dependence on $\ptjthr$: FtW corrections grow
from 6\% up to more than 40\%, and the $K$-factor decreases very fast due to
the presence of a soft singularity at $\ptjthr\to 0$.  For a jet veto with
$\ptjthr=30\,\GeV$ we observe a 98\% suppression of the $\wwbb$ cross
section.  Yet the moderate size of the $K$-factor and NLO scale variations
indicates that the perturbative expansion is still rather stable in this
regime.  In the 1-jet bin, the limit of small $\ptjthr$ is driven by the effect of the
veto on the second jet, and NLO and FtW corrections behave rather similarly as
for the 0-jet bin in this region.  In the opposite regime, $\ptjthr$ mainly
acts as a lower $p_{\rT}$ bound for the first jet, and $\ttbar$ production
with LO on-shell kinematics turns out to be kinematically disfavoured at
large $\ptjthr$, while the relative importance of NLO jet emission and FtW
effects increases quite dramatically.

\begin{figure*}
\begin{center}
{\includegraphics[width=.45\textwidth]{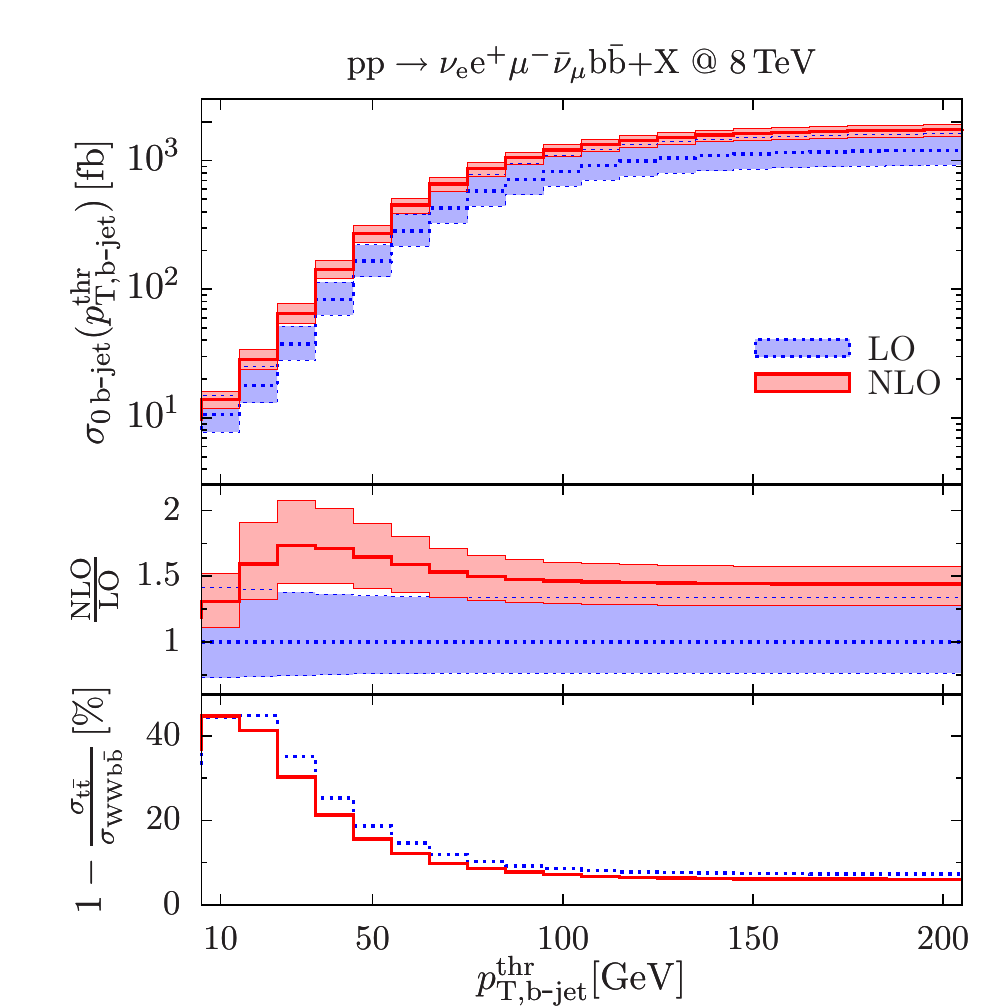}}
\hspace{10mm}
{\includegraphics[width=.45\textwidth]{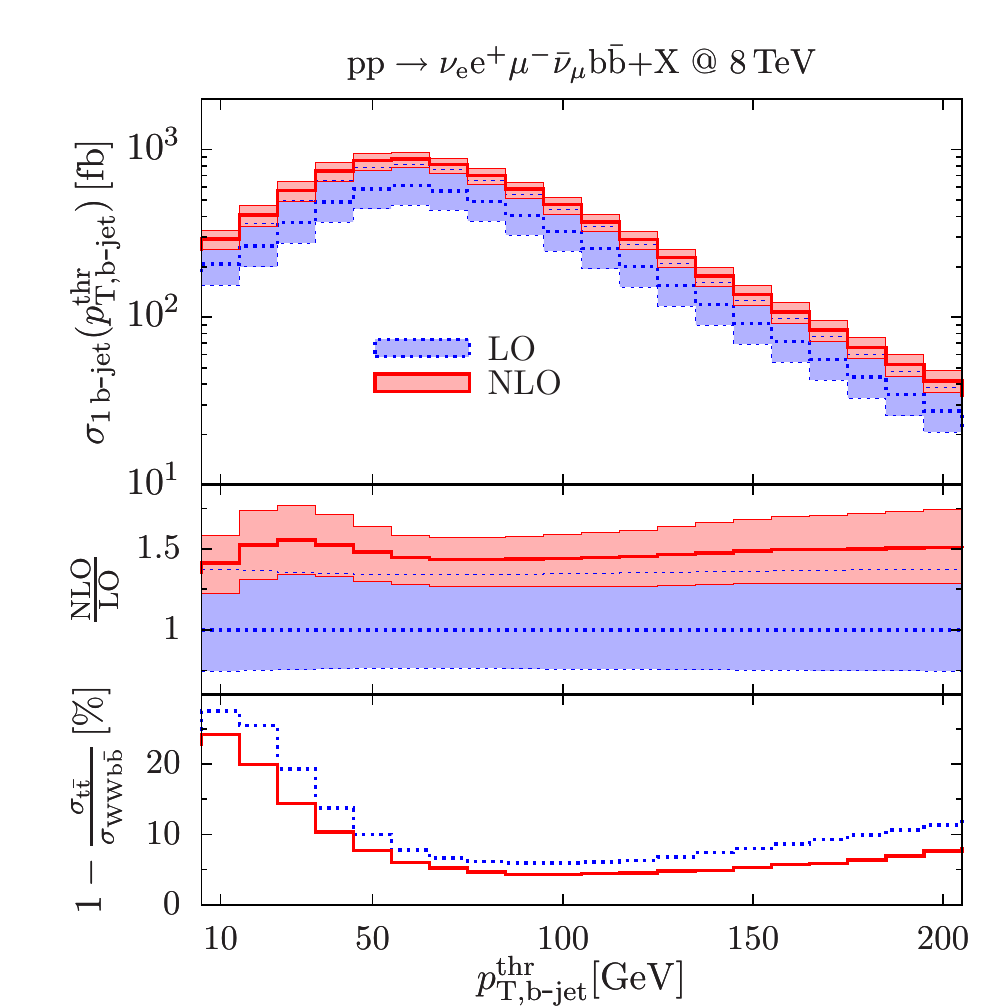}}
\end{center}
\caption{
LO and NLO $\wwbb$ cross sections in the exclusive bins with $\nb=0$ (left) and $\nb=1$ (right) 
b-jets versus the b-jet-$p_\rT$ threshold. Same conventions as in \reffi{fig:ptjet}.}
\label{fig:ptbjet}
\vspace{-1.mm}
\end{figure*}

Analogous results for exclusive bins with $\nb=0$ and $\nb=1$ b-jets are displayed in
\reffi{fig:ptbjet}. In this case the reduced sensitivity of b-jet bins to
NLO real emission is clearly reflected in the much better stability of the $K$-factor
with respect to variations of $\ptbjthr$. Similarly as for jet bins, FtW corrections 
are strongly enhanced at small $p_{\rT}$.
This effect can be attributed to the single-top Wt channels, 
and the inclusion of $\ttbar$--$\wt$ interferences, as in the present $\wwbb$ calculation,
is clearly advisable in this regime.

\begin{figure*}
\begin{center}
{\includegraphics[width=.45\textwidth]{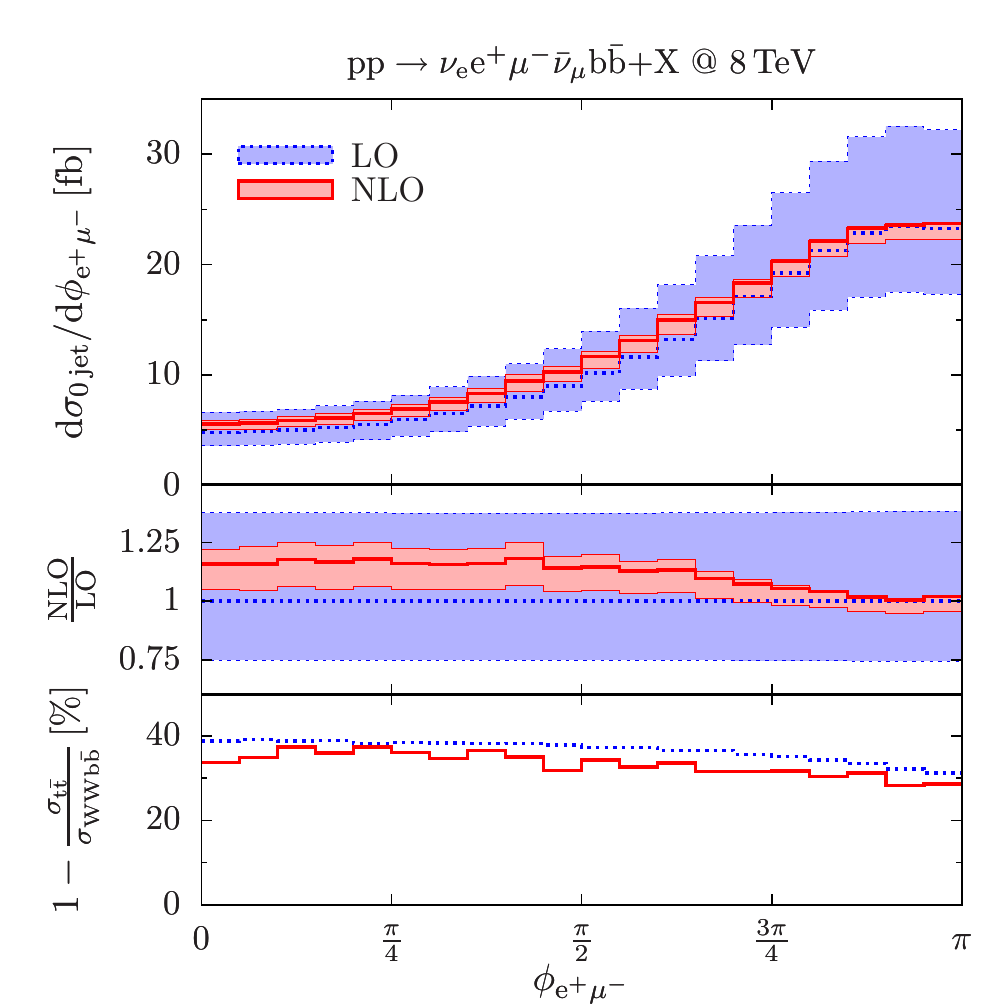}}
\hspace{10mm}
{\includegraphics[width=.45\textwidth]{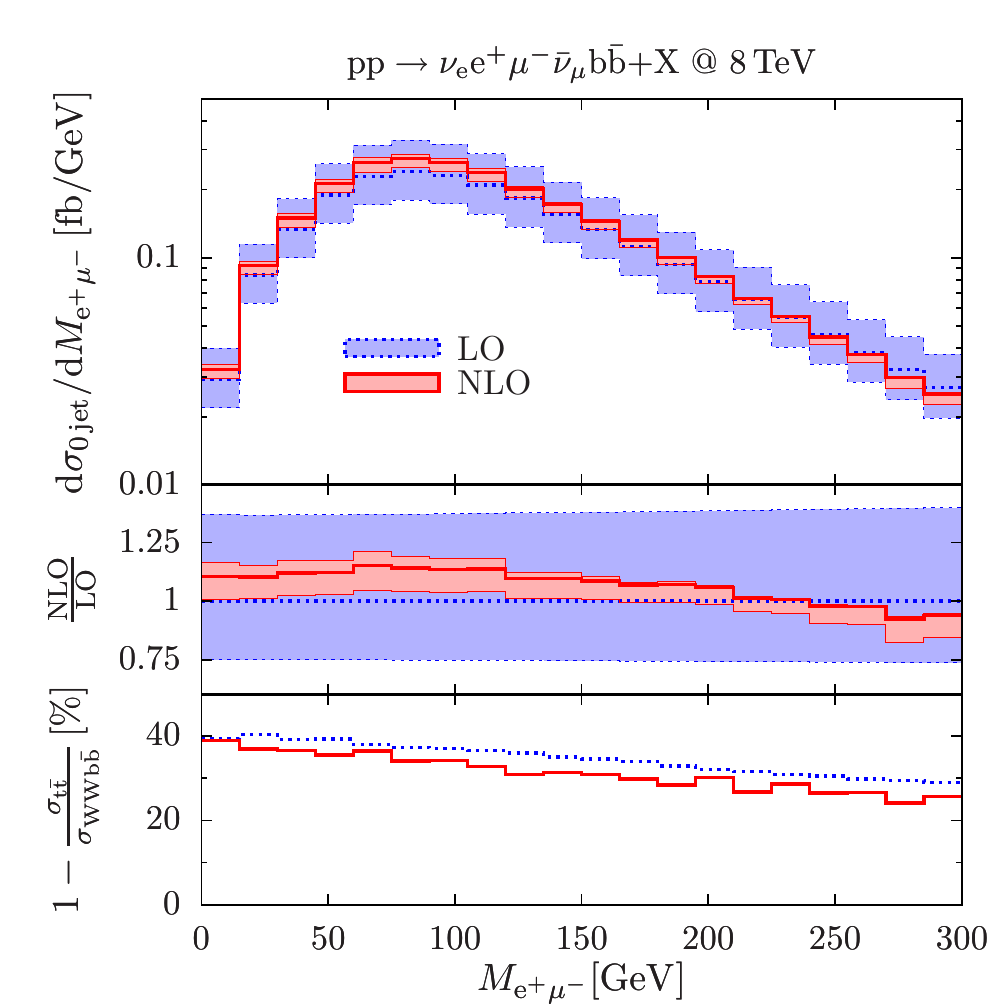}}
\end{center}
\caption{Differential distributions in the 0-jet bin: azimuthal-angle separation (left) 
and invariant mass (right) of the two charged
leptons.
Same conventions as in \reffi{fig:ptjet}.
}
\label{fig:pt0jet}
\vspace{-1.mm}
\end{figure*}

Finally, in \reffi{fig:pt0jet} we show distributions in the azimuthal-angle-separation 
and in the invariant mass of charged leptons in the 0-jet bin. 
These observables play a key role for the measurement of the $\PH\to\PWp\PWm$ signal 
at the LHC, and the accurate  modelling of top-backgrounds is very important
for the experimental analyses. In this context, \reffi{fig:pt0jet} shows
that NLO and FtW effects are quite significant.
In particular, the impact of FtW contributions reaches up to 40\%.
Shape distortions due to the kinematic dependence of FtW and NLO contributions 
are at the 10\% level, and scale variations do not exceed 10\% at NLO. 
The fact that FtW corrections are fairly stable with respect to NLO
corrections provides further evidence of the stability 
of the perturbative description.

\section{Summary and conclusions}
We have presented a complete NLO simulation of $\wwbb$ production at
the LHC, including W-boson decays in the opposite-flavour di-lepton channel, finite W- and
top-width effects, and massive b-quarks in 4F scheme.  The finite b-quark
mass acts as a regulator of collinear singularities and allows one to
describe the full b-quark phase space, including single-top contributions
that arise from initial-state $\Pg\to\bbbar$ splittings followed by $\Pg\Pb\to
\PW\Pt$ scattering.  This yields a gauge-invariant description
of top-pair, single-top, and non-resonant $\wwbb$ production
including all interferences at NLO QCD.  We
introduced a dynamical scale choice aimed at an improved perturbative
stability of initial-state $\Pg\to\bbbar$ splittings in single-top
contributions.  Using this scale, the NLO $\wwbb$ cross section in bins with
0, 1 and 2 jets features NLO scale uncertainties at the 10--15\% level.  The
more conventional choice $\mu_0=\Mt$ yields similarly small NLO uncertainties 
in jet bins. While providing further evidence of the 
good convergence of the perturbative expansion, this means 
that a sophisticated dynamical scale is unnecessary for the
rather inclusive observables considered in this letter. However, such a dynamical scale
might become important for more exclusive observables, like jet-$p_\rT$ distributions.

Finite-top-width corrections
mainly originate from \linebreak single-top and off-shell $\ttbar$ contributions. They
represent 6\% of the integrated cross section and are strongly sensitive to the jet
multiplicity.  In the 2-jet bin they are as small as 2\%,
while in the 1- and 0-jet bins they reach the 16\% and 32\% level, respectively.  Also 
NLO corrections vary quite strongly with the jet multiplicity.
Moreover, finite-top-width contributions receive quite different corrections
as compared to on-shell $\ttbar$ production.

The non-trivial interplay of NLO and finite-width effects 
is especially relevant for the 0- and 1-jet bins. It plays an important role for the
accurate description of associated $\PW\Pt$ production, as well as for
top-backgrounds to $\PH\to\PWp\PWm$ and to other searches based on leptons,
large missing energy and jet vetoes.
All employed tools are fully automated and can be easily exploited
to extend the present results to the like-flavour di-lepton channel 
or to simulate any other Standard-Model process at NLO QCD.

\begin{acknowledgements}

We thank A.~Denner, S.~Dittmaier and L.~Hofer for providing us with the
one-loop tensor-integral library \Collier.  We are grateful to
S.~H\"oche and F.~Siegert for \Sherpa technical support.
Our research
is funded by the SNSF  and supported, in part, by the European
Commission through the
network PITN-GA-2010-264564 ({\it LHCPhenoNet}).

\end{acknowledgements}

\bibliography{wwbb4f}

\end{document}